\documentclass{aa}
\usepackage{graphicx}
\usepackage{natbib}
\usepackage[varg]{txfonts}

\newcommand{\HI}{{\ion{H}{1}}}

\newcommand{\kms}{$\,$km$\,$s$^{-1}$}
\newcommand{\ergs}{$\,$erg$\,$s$^{-1}$}

\newcommand{\mJybeam}{mJy beam$^{-1}$}
\newcommand{\muJybeam}{$\mu$Jy beam$^{-1}$}
\newcommand{\msun}{{$M_\odot$}}
\newcommand{\msunyr}{{$M_\odot$ yr$^{-1}$}}
 
\newcommand{\miriad}{{{MIRIAD}}}
\newcommand{\casa}{{{CASA}}}

\def\HI{H{\,\small I}}
\def\HI{\ion{H}{i}}

\newcommand{\ltsima} {$\; \buildrel < \over \sim \;$}
\newcommand{\gtsima} {$\; \buildrel > \over \sim \;$}
\newcommand{\lta} {\lower.5ex\hbox{\ltsima}}
\newcommand{\gta} {\lower.5ex\hbox{\gtsima}}

\begin{document}

\title{The fast molecular outflow in the Seyfert galaxy IC~5063\\ as seen by ALMA}
\titlerunning{ The fast molecular outflow in the Seyfert galaxy IC~5063 as seen by ALMA}
\authorrunning{Morganti et al.}
\author{Raffaella Morganti\inst{1,2}, Tom Oosterloo\inst{1,2}, J. B. R.  Oonk\inst{1,3}, Wilfred 
Frieswijk\inst{1},  Clive Tadhunter \inst{4}}
\institute{ASTRON, the Netherlands Institute for Radio Astronomy, Postbus 2, 7990 AA, Dwingeloo, The 
Netherlands.
\and
Kapteyn Astronomical Institute, University of Groningen, P.O. Box 800,
9700 AV Groningen, The Netherlands
\and
Leiden Observatory, Leiden University, P.O. Box 9513, 2300 RA Leiden, The Netherlands
\and
Department of Physics and Astronomy, University of Sheffield, Sheffield, S7 3RH, United Kingdom
}
\offprints{morganti@astron.nl}

\date{Received ...; accepted ...}

\date{\today}

\abstract{We use high-resolution (0.5 arcsec) CO(2-1) observations performed
  with the Atacama Large Millimetre/submillimetre Array to trace the
  kinematics of the molecular gas in the Seyfert 2 galaxy \object{IC
    5063}. The data reveal that the kinematics of the gas is very complex. A
  fast outflow of molecular gas extends along the entire radio jet ($\sim 1$
  kpc), with the highest outflow velocities about 0.5~kpc from the nucleus, at
  the location of the brighter hot-spot in the W lobe. The ALMA data show that
  a massive, fast outflow with velocities up to $ 650$ \kms\ of cold molecular
  gas is present, in addition to one detected earlier in warm H$_2$, \HI\ and
  ionised gas. All phases of the gas outflow show similar kinematics. IC~5063
  appears to be one of the best examples of the multi-phase nature of
  AGN-driven outflows.
  \\
  Both the central AGN and the radio jet could energetically drive the
  outflow. However, the characteristics of the outflowing gas point to the
  radio jet being the main driver. This is an important result, because
  IC~5063, although one of the most powerful Seyfert galaxies, is a relatively
  weak radio source ($P_{\rm 1.4~GHz} = 3 \times 10^{23}$ W Hz$^{-1}$).
  \\
  All the observed characteristics can be described by a scenario of a radio
  plasma jet expanding into a clumpy medium, interacting directly with the
  clouds and inflating a cocoon that drives a lateral outflow into the
  interstellar medium. This model is consistent with results obtained by
  recent simulations such as those of Wagner et al.. A stronger, direct
  interaction between the jet and a gas cloud is present at the location of
  the brighter W lobe. This interaction may also be responsible for the
  asymmetry in the radio brightness of the two lobes.
  \\
  Even assuming the most conservative values for the conversion factor
  CO-to-H$_2$, we find that the mass of the outflowing gas is between $1.9$
  and $4.8 \times 10^7$ \msun, of which between $0.5$ and $1.3 \times 10^7$
  \msun\ is associated with the fast outflow at the location of the W lobe.
  These amounts are much larger than those of the outflow of warm gas
  (molecular and ionized) and somewhat larger than of the \HI\ outflow. This
  suggests that most of the observed cold molecular outflow is due to fast
  cooling after being shocked. This gas is the end product of the cooling
  process, although some of it could be the result of only partly
  dissociated clouds. Our CO observations demonstrate that fast outflows of
  substantial masses of molecular gas can be driven by relativistic jets,
  although in the case of IC~5063 the outflows are not fast enough to remove
  significant amounts of gas from the galaxy and the effects are limited to
  the central $\sim 0.5$ kpc from the centre. }

\maketitle  

\section{Introduction}
\label{sec:introduction}

The discovery of the large variety of structures and phenomena traced by cold
gas in early-type galaxies has made these objects much more complex than
historically perceived. Although not a complete surprise (see e.g. the work of
\citet{vanDriel1991} and \citet{Wiklind1995}, the plethora of new results,
made possible by the technical improvements in the available and new observing
facilities, has opened new and important avenues for the study of formation
and evolution of early-type galaxies using this component of their
inter-stellar medium (ISM) \citep[see,
e.g.,][]{Young2011,Serra2012,Davis2013,Alatalo2013,Serra2014}.
 
 
\begin{figure*}
\centering
\includegraphics[width=12cm,angle=0]{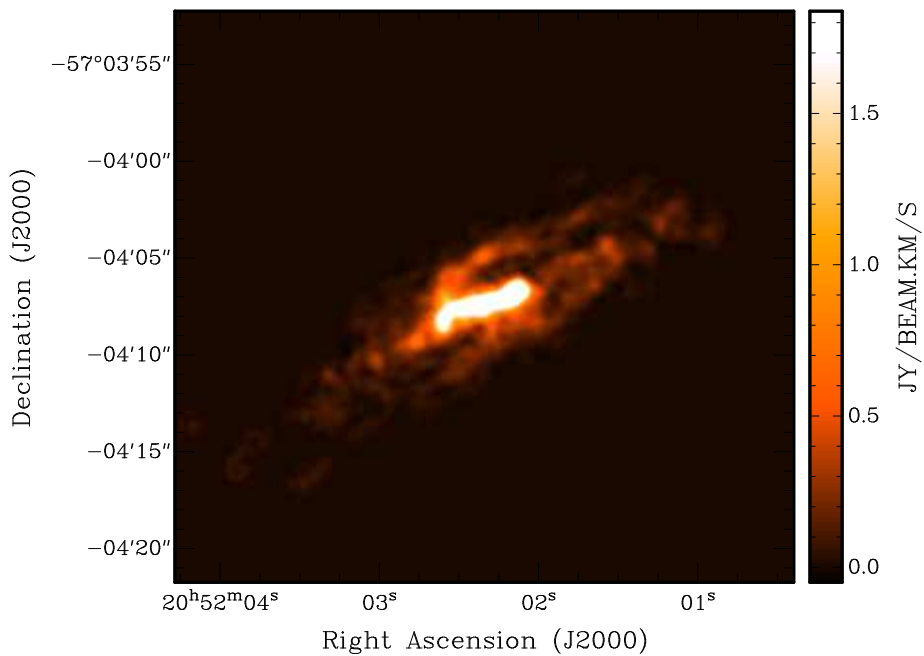}
\caption{Total intensity image representing the distribution of the CO(2-1) in
  IC 5063 and showing the striking brightness contrast between the inner,
  bright CO and the fainter outer disk.  The details of the bright inner
  region can be seen in Fig.\ \ref{fig:ContCO}.  The resolution of this image
  is $0.54 \times 0.45$ arcsec (PA $= -31.7^{\circ}$). This image is not
  corrected for the primary beam of ALMA.}
\label{fig:TotInt}
\end{figure*}


In addition to being an effective tracer of the formation history of galaxies,
the cold gas has revealed unexpected characteristics that have allowed to use
it also for tracing extreme energetic phenomena like massive and fast
AGN-driven gas outflows \citep[][and refs therein]
{Morganti2005a,Morganti2005b,Feruglio2010,Dasyra2012,Morganti2013a,
  Morganti2013b,Mahony2013,Morganti2014}.  This finding has challenged our
ideas of how the energy released by an AGN may interact with its surroundings
\citep[see, e.g.,][for reviews on the
topic]{Tadhunter2008,Fabian2013,Combes2014}. Cold gas has been found to
represent the dominant mass component of the outflows observed \citep[see,
e.g.,][for some examples]
{Morganti2005a,Morganti2005b,Kanekar2008,Holt2009,Feruglio2010,Alatalo2011,
  Dasyra2011,Dasyra2012,Morganti2013a,Morganti2013b,Cicone2014,Garcia2014}.
This makes them particularly interesting for their potential impact on the ISM
of the host galaxy. In the effort of understanding the mechanisms that can
regulate the black-hole growth and the quenching of star-formation in
early-type galaxies, the discovery, in a growing number of objects, of these
AGN-driven massive outflows represents an important step toward building a
more complete and realistic view of how feedback may work. Thanks to detailed
studies of these phenomena in nearby objects, better constraints (i.e.\
distribution and kinematics of the outflowing gas, energy injected into the
ISM, etc.) can be provided to theoretical models of galaxy formation that now
routinely include feedback effects.

Different mechanisms have been proposed to accelerate the gas. Although (broad)
winds from the accretion disk, launched by radiation pressure, interacting and
shocking the surrounding medium is often favoured
\citep[e.g.][]{Zubovas2012,Zubovas2014,Costa2014,Cicone2014}, the role of the
radio plasma has also gained interest. Indeed, this mechanism can be
particularly relevant in early-type galaxies where up to 30\% of the high-mass
galaxies are radio loud \citep{Best2005}. A number of studies \citep{Birzan2008,Cavagnolo2010,McNamara2012} have shown  an high efficiency of the
coupling between the radio plasma and the surrounding ISM/IGM, and 
mechanical power of the radio jets exceeding the synchrotron power.
These results  suggest that this mechanism
can be relevant also for relatively weak radio sources \citep[see, e.g., NGC
1266 and NGC 1433,][respectively]{Alatalo2011,Combes2013}.

Expanding the number of known outflows of atomic (\HI) and molecular gas in
order to provide better statistics is the focus of a number of recent studies
\citep[see, e.g.,][]{Gereb2014a,Gereb2014b,Cicone2014}. However, the other
important direction where progress needs to be made is to spatially resolve
the outflows and image in detail the distribution, kinematics and the physical
properties of the gas. In the case of 4C~12.50, VLBI observations have, for
the first time, allowed this to be established for the \HI\ component
\citep{Morganti2013a}. For the warm molecular gas, the most spectacular and
convincing example of a jet-driven outflow has been found in IC~5063 by
\citet{Tadhunter2014}. In this object, the warm component of the molecular gas
has been identified at the location of the interaction between the radio jet
and the ISM, 0.5 kpc from the central AGN, where an \HI\ outflow was also
found \citep{Oosterloo2000}. AGN-driven outflows of cold molecular gas have
been so far only barely resolved \citep[e.g.][]{Feruglio2010,Cicone2014} not
always allowing a detailed modelling of the outflowing gas and its conditions.
The possible exceptions are M51 \citep{Matsushita2007,Matsushita2015} and
NGC~4258 \citep{Martin1989,Krause2007}. The situation is now changing with the
advent of ALMA. The detailed case of NGC~1068 \citep{Garcia2014} is an example
of the dramatic improvement that ALMA allows to make.

All together, these studies are the first to provide details on the physical
parameters and the conditions of the gas in the outflows. One of these
objects, the Seyfert 2 galaxy IC~5063, is the target of the ALMA observations
presented here showing, for the first time, the complexity of the kinematics
of the cold gas under the effect of the radio jets.

\section{A case study: IC~5063}

IC 5063 is a Seyfert 2 object hosted by an early-type galaxy with a prominent
dust lane which is seen edge on. Together with NGC~1068, it is one of the most
radio bright Seyfert objects with a power\footnote{Assuming a Hubble constant
  $H_{\circ}$= 70 km s$^{-1}$ Mpc$^{-1}$ and $\Omega_\Lambda=0.7$ and
  $\Omega_{\rm M} = 0.3$, we use an angular scale distance to IC 5063 of 47.9
  Mpc, implying a scale of 1 arcsec = 232 pc.} of
$P_{\rm 1.4GHz} = 3 \times 10^{23}$ W Hz$^{-1}$ \citep{Tadhunter2014}. In the
radio continuum, IC~5063 shows a triple structure of about 4 arcsec in size
(about 0.93 kpc; see \citet{Morganti1998}) aligned with the dust-lane. \HI\
emission ($8.4 \times10^9$ M$_\odot$) is traced in a regularly rotating warped
disk with radius $\sim 2$ arcmin (about 28 kpc) and which is associated with
the system of dust lanes. The likely counterpart of the \HI\ disk is also
detected in CO as originally observed with SEST in CO(1-0) by
\citet{Wiklind1995} and more recently with APEX in CO(2-1) by
\citet{Morganti2013b}.

IC~5063 was the first object where a fast, AGN driven massive outflow of \HI\
was detected \citep{Morganti1998}. This outflow was found to be located
against the brightest radio hot-spot, about 0.5 kpc from the radio core
\citep{Oosterloo2000} and this has been interpreted as evidence that the
outflow is driven by the interaction of the radio jet with the surrounding ISM
where the jet is moving in the plane of the large-scale gas disk of IC 5063.
This hypothesis has been further strengthened by observations (using ISAAC on
the VLT) of warm molecular gas (H$_2$ at 2.2 $\mu$m) which have, for the first
time, shown the presence of outflowing molecular gas co-spatial with the
bright radio hot-spot \citep{Tadhunter2014}. A signature of outflow of cold
molecular gas has been found using CO(2-1) observations with APEX
\citep{Morganti2013b}. The limited spatial resolution of these observations
did not allow to determine where the outflow is occurring in relation to the
radio structure and the main goal of the ALMA data presented in this paper is
to remedy this situation. The data presented here complement the information
about ionized \citep{Morganti2007}, warm molecular \citep{Tadhunter2014} and
cold atomic \citep{Morganti1998,Oosterloo2000}, making IC~5063 the best case
of a multi-phase AGN-driven outflow studied so far.

The sensitivity and spatial resolution of the new ALMA observations allow us
to show for the first time the detailed distribution and kinematics of the
cold molecular fast outflow, providing a major step forward in the
understanding of the interaction between the energy released by the AGN and
the surrounding medium.


\begin{figure}
\centering
\includegraphics[width=\hsize, keepaspectratio]{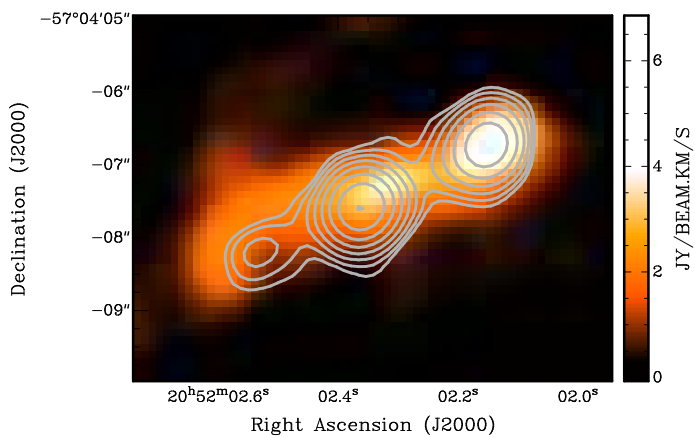}
\caption{Contours of the 230~GHz continuum emission superposed to the central
  region of the total intensity of the CO(2-1) illustrating the spatial
  correlation between the two. Contour levels are 90, 180, 360, 720, 1440,
  2880, 5760 and 15120 \muJybeam.}
\label{fig:ContCO}
\end{figure}


\section{The ALMA data}

IC~5063 was observed during Cycle 1 on the 27 May 2014 with the Atacama Large
Millimeter/submillimeter Array (ALMA) using Band 6, covering simultaneously
the CO(2-1) centered on 227.968~GHz and two additional base-bands (at 232 and
247~GHz) for imaging of the continuum.  The observations were pointed at the
nucleus of IC~5063, with a single-pointing field of view (FoV) of
$\sim 27$\arcsec\ and making use of the correlator in Frequency Division Mode
with a 1.875 GHz bandwidth (corresponding to 2466 \kms\ at the frequency of
CO(2-1)).  The frequency resolution was 976.5 kHz (1.28 \kms), but channels
were combined when making the image cube to provide a better match to the
observed line widths. The observation made use of 31 antennas with a maximum
baseline of 1.5 km. The duration of the observations was 1 h which was
required to reach the sensitivity to locate and trace the outflow of molecular
gas. The initial calibration was done in
\casa\footnote{http://casa.nrao.edu/}\ \citep[v4.2][]{McMullin2007} using the
ALMA reduction scripts. These calibrated $uv$ data were subsequently exported
to \miriad\footnote{http://www.atnf.csiro.au/computing/software/miriad/}
\citep{Sault1995} which was used to perform additional self-calibration which
improved the quality of the images significantly. All further reduction steps
(continuum subtraction, mapping/cleaning) were also done in \miriad.

The data cubes were made using various Briggs weightings in order to explore
the optimum for imaging. However, the images did not show any major
differences (owing to the good $uv$ coverage of ALMA). Here we present the
final data cubes obtained using Briggs robust = 0. The line cube was produced
using a velocity resolution, after Hanning smoothing, of 20 \kms. At this
velocity resolution, the r.m.s.\ noise per channel is $\sim 0.3$ \mJybeam.
The large array configuration used is providing a resolution of the line data
of $0.54 \times 0.45$ arcsec (PA $= -31.7^{\circ}$), nicely allowing the line
emission to be resolved in relation to the radio continuum emission.

In Fig. \ref{fig:TotInt} the total intensity of the CO(2-1) is shown. The
single pointing ALMA observations detects molecular gas up to a radius of
$\sim 10^{\prime\prime}$, covering therefore only the very inner part ($r < 2$
kpc) of the \HI\ disk observed with ATCA \citep{Morganti1998}.  The bulk of
the emission covers the same range of velocities (between 3140 to 3630 \kms)
as traced by the APEX observations \citep{Morganti2013b}. However, the wider
band of ALMA allows to trace a much broader component of blueshifted gas,
going down to velocities of $\sim 2750$ \kms, i.e.\ up to $\sim 650$ \kms\
blueshifted compared to the systemic velocity.

A continuum image was also produced from the 2 additional basebands using
uniform weighting. The r.m.s.\ noise of the continuum image is 30
\muJybeam. The restoring beam $0.45 \times 0.42$ arcsec (PA$=
-30.4^{\circ}$).
Figure \ref{fig:ContCO} shows the comparison between the mm continuum and the
molecular gas in the inner region.

Finally, the CS(5-4) line at 242.205 GHz was included in a fourth base-band,
but no emission from this spectral line was detected. The noise level in this
cube, not quite uniform across the band, ranges between 0.3 and 0.4 \mJybeam.


\begin{figure}
\centering
\includegraphics[width=\hsize, keepaspectratio]{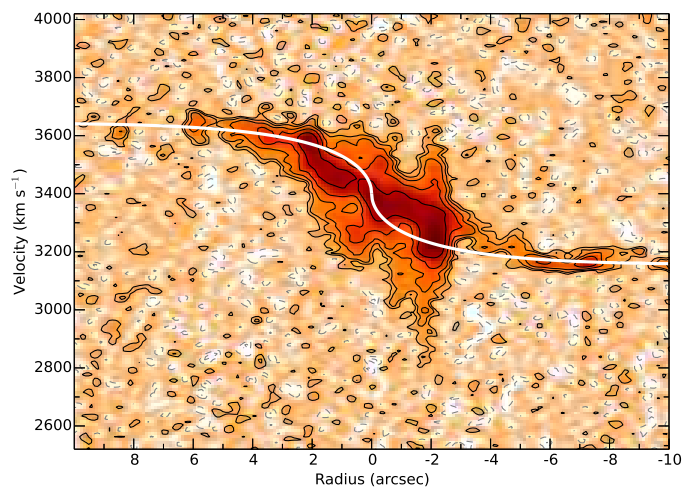}
\caption{Integrated position-velocity map taken along the major axis of IC
  5063 where the data were integrated over a length of 2 arcsec in the
  direction perpendicular to the major axis. The white line gives the rotation
  curve we derived from the photometry of \citet{Kulkarni1998} and indicates
  the expected kinematics of gas following regular rotation. Contour levels
  are --3, --1.5, 1.5, 3, 6, 12 and 24 $\sigma$. Positive radii are  east of the centre.}
\label{modelRotCurve}
\end{figure}


\section{Results}

The main result of our observations is the detailed view of the complexity of
the distribution and kinematics of the gas in the central regions.  The data
clearly demonstrate that, apart from the regularly rotating large-scale disk,
the distribution and kinematics of the gas in the inner regions is strongly
affected by the radio jet.

Figure \ref{fig:TotInt} shows that the distribution of the CO(2-1) follows two
structures: a high-surface-brightness inner structure of about the size of the
radio continuum and a lower-surface-brightness, larger disk.  This large disk
is likely the molecular counterpart of the \HI\ disk known to be present in
this galaxy (see Sec. \ref{sec:regular}).  Some of the gas in this disk
appears to be distributed along arm-like features. However, the relatively
edge-on orientation of IC 5063 makes it difficult to identify and characterise
these structures well.

The bright and elongated structure observed in the central regions shows a
striking correspondence with the radio continuum, providing the first clear
signature of the impact of the radio plasma on the ISM. Figure
\ref{fig:ContCO} illustrates how the CO bright emission wraps around the
radio continuum, with the brightest CO component coincident with the brighter
radio lobe on the western side.

Below we describe kinematics and other properties of these CO components.

\subsection{A complex, extended outflow of molecular gas }

The kinematics of the gas provides a striking view of what is
happening in the central region.  In Fig.\ \ref{modelRotCurve} we show a
position-velocity ($pv$) diagram of the CO gas taken along the major axis of
the galaxy (which coincides, to within a few degrees, with the orientation of
the radio continuum source). Given the relatively edge-on orientation of the
system, a position-velocity diagram gives a better view of the kinematics of
the gas compared to a velocity field. To make this $pv$ diagram, we have
integrated the data over a length of 2\arcsec\ in the direction perpendicular
to the major axis. Therefore this map shows all the gas in the vicinity of the
radio source, but some of the gas associated with the large-scale disk is not
included. To indicate what can be expected for the kinematics of regularly
rotating gas, we have overplotted a rotation curve based on the HST photometry
of IC 5063 of \citet{Kulkarni1998} (see below).


\begin{figure}
\centering
\includegraphics[width=\hsize, keepaspectratio]{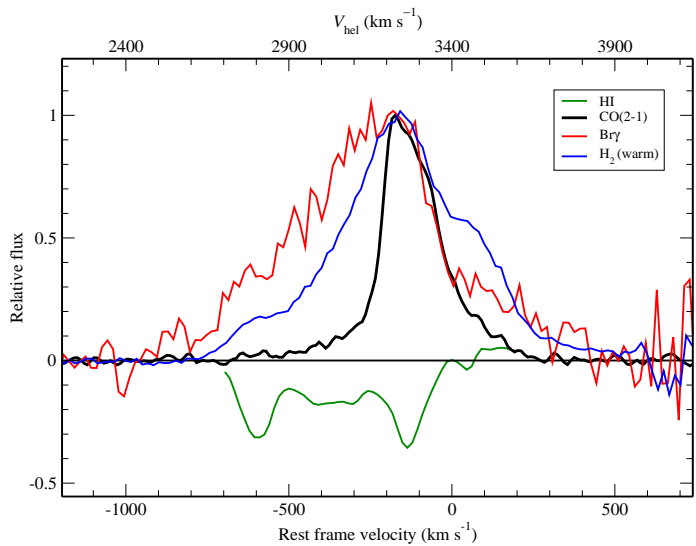}
\caption{Normalized spectra of H$_2$ 1-0 S(1) (black solid line) and
  Brackett-gamma \citep{Tadhunter2014} and CO (this work) integrated over
  the western radio lobe of IC~5053 are compared with the spatially integrated
  \HI\ 21-cm absorption (from \citealt{Morganti1998}). The plot shows that all 
  the components have gas reaching velocities of $ -650$ \kms\ or beyond.  }
\label{fig:Width_outflow}
\end{figure}


Figure \ref{modelRotCurve} clearly illustrates the strikingly complex
kinematics of the gas in the region co-spatial with the radio plasma. Beyond a
radius of $2^{\prime\prime}$ (i.e. $\sim 460$ pc), the gas follows the
expected rotation pattern fairly well and this gas corresponds to the
large-scale disk, unaffected by the radio plasma.  If gas with such complex
kinematics would have been present outside the region of the radio emission,
the sensitivity of our observations would be able to see. Instead, the change
between complex to regular kinematics is quite abrupt.  However, inside this
radius, most of the gas has velocities quite different from what is expected
based on the rotation curve. In fact, a significant fraction of the gas is
found in the so-called "forbidden quadrants" of the $pv$ diagram, i.e.\ those
parts where co-rotating gas cannot appear.  Examples of this are the gas at
$r \sim -2^{\prime\prime}$ and with velocities around 3500 \kms \ and,
similarly, the gas at $r\sim1^{\prime\prime}$ and $V \sim 3200$-3300 \kms. In
addition, for $-2^{\prime\prime} < r < 0^{\prime\prime} $ there is a
substantial amount gas with velocities that are well outside the range of
expected velocities, indicating that the kinematics of this gas is not
determined by gravity alone. This gas is most likely related to the fast,
outflowing \HI\ gas that was discovered in absorption by \citet{Morganti1998}.
As we will show below (Sec.\ \ref{sec:modeling}), the distribution and
kinematics of the molecular gas with anomalous velocities unequivocally point
to the presence of outflowing gas and can be understood in terms of a lateral
outflow driven by the radio jet.

In further agreement with earlier observations, also of other gas phases
\citep{Morganti2007,Tadhunter2014,Morganti1998,Oosterloo2000}, the most
extreme velocities are confirmed to be present at the location of the bright W
radio lobe, about 0.5~kpc from the nucleus. This, again, is a strong
indication that the most likely origin of these extreme kinematics must be the
jet. The most extreme velocities of the anomalous gas are blueshifted by more
than 500 \kms\ and redshifted by more than 400 \kms\ w.r.t.\ the regularly
rotating gas at that location.  The velocity profiles of the different gas
components integrated over the region of the western hot-spot (except for the
\HI\ that shows the profile integrated over the full radio continuum region
although the most blue shifted HI absorption is most likely to occur only
against the continuum of the bright W hot spot) are compared in
Fig. \ref{fig:Width_outflow}.  In all cases (\HI\ absorption, the ionized gas
and the warm and cold molecular gas), the blueshifted component of the gas
reaches velocities of $\sim -700$ \kms
\citep[][respectively]{Morganti1998,Morganti2007,Tadhunter2014}.

The rotation velocity of IC 5063 is about 240 \kms\ which means that the
escape velocity in the inner regions is likely to be above 550 \kms\
\citep{Binney2008}.  We observe basically no gas with such extreme
  velocities (see Fig.\ \ref{modelRotCurve}). This means that, allowing for
  projection effects, at most a very small fraction of the anomalous gas will
leave IC 5063 and that any possible effect on future star formation will have
to be through an increased energy input in the ISM and not through gas
removal.  A similar situation has been observed to be the case in NGC 1266
  \citep{Alatalo2015}.

\subsection{The bright inner CO(2-1) distribution}
\label{sec:BrightInnerCO}

A characteristic of the CO(2-1) distribution that clearly appears from
Fig.\ \ref{fig:TotInt}, as well as from the position-velocity diagram in
Fig.\ \ref{modelRotCurve} is that, although the large-scale, regular rotating
disk is clearly seen, the emission from the region around the radio jet is
much brighter in CO(2-1). Two possibilities can be suggested to explain this.

The bright inner CO (J=2-1) emission in IC~5063 can be excited by a variety of
physical processes, such as shocks \citep[e.g.][]{Hollenbach1989}, cosmic rays
\citep[e.g.][]{Sternberg1989,Ferland2009,Meijerink2011,Bayet2011}, mechanical
heating \citep{Loenen2008,Meijerink2011} and photodissociation (including
  X-rays; e.g.\ \citealt{Tielens1985,Maloney1996, Meijerink2007}). Given that
we have measured only one molecular line (CO J=2-1) at sufficient spatial
resolution to resolve the molecular gas in this region we not able to
discriminate between these different processes.

The difference between the inner and outer regions may reflect differences in
the temperature and/or excitation of the molecular gas in the two regions due
to the influence of the jet/cloud interaction. As discussed by
\citet{Bolatto2013} for starbursts and ULIRGs, to first order, higher gas
temperatures yield brighter CO emission, decreasing $X_{\rm CO}$ for the
regions affected by the interaction. However, the increase in density may
partly compensate for this effect, as pointed out by
\citet{Bolatto2013}. Therefore a certain level of compensation may occur,
lessening the impact of environment on the conversion factor.  Without a
detailed study of the physical conditions of the molecular gas, e.g.,
using the full CO ladder, it is not possible to fully quantify these effects.

On the other hand, the kinematics of the gas is clearly indicating that the
effect of the radio jet is relevant for the distribution of the gas in the
inner regions. The cocoon of shocked material created by the jet moving
through the ISM may be able to push the gas and move it outward.   
This explanation would be consistent with what is proposed by
theoretical models \citep[e.g.][]{Wagner2011} and we will consider such a
scenario in more detail in Sec. \ref{sec:modeling}.

This scenario also implies that shocks will likely play a prominent role in
the heating and excitation of the molecular gas. In addition the outward
moving jet and associated shocks will boost the ionizing flux of cosmic rays
(CR). Several theoretical studies have shown that high CR rates can
significantly modify the chemistry and conditions of the molecular gas
\citep[e.g.][]{Ferland2009,Bayet2011,Meijerink2011}.

\subsection{The regularly rotating CO disk}
\label{sec:regular}

Outside the region of the radio continuum emission, the CO(2-1) shows regular
kinematics. The characteristics are of a rotating disk with the western side
approaching. This structure most likely represents the molecular counterpart
of the regular disk already observed in \HI\ and in ionized gas
\citep{Colina1991,Morganti1998,Morganti2007}.  Indeed, the rotation velocities
at large radii of this molecular gas are very similar to those of the \HI,
with an amplitude of the rotation of $\sim$246 \kms.  However, given the
limited field-of-view of the primary beam of ALMA, we cannot determine the
full extent of the molecular component of the large-scale disk. Furthermore,
the comparison with the velocity profiles of the \HI\ in the inner regions is
limited by the low spatial resolution of the available \HI\ observations.
Large disks and rings of regularly rotating molecular gas are now known to be
relatively common structures in early-type galaxies and are observed in more
than 50\% of CO detected early-type galaxies \citep{Alatalo2013}. Thus, in
this respect, IC~5063 is not an exception.

We have modelled the rotation of the regular CO distribution using the HST
photometry of \citet{Kulkarni1998}. This photometry, in the bands F110W, F160W
and F222M, shows that the unobscured light distribution of the inner regions
of IC~5063 can accurately be described using a de Vaucouleurs' law.  Figure
\ref{modelRotCurve} shows, overplotted on a $pv$ diagram, the rotation curve
expected for the observed light distribution where we have assumed that the
light traces the mass distribution and that the mass distribution is
spherical. The horizontal scale of the rotation curve is set by the observed
effective radius (21\farcs{5} corresponding to 4.3 kpc) while we have set the
amplitude of the rotation curve to match the rotation velocities of the very
outer regions in Fig.\ \ref{modelRotCurve}.


\begin{figure}
\centering
\includegraphics[width=\hsize, keepaspectratio]{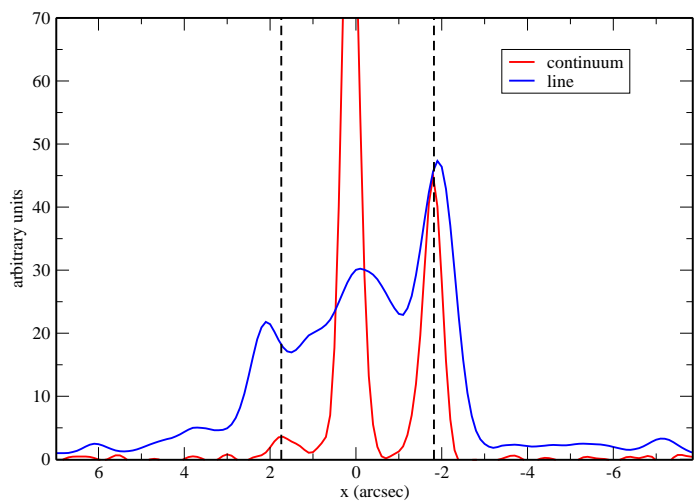}
\caption{Brightness distribution, in arbitrary units, along the
  direction of the radio jet of the synchrotron continuum emission (red line)
  and of the CO(2-1) emission (blue line). The dashed vertical lines indicate
  the peaks of the continuum hot spots, showing that the bright CO slightly
  extends beyond the hot spots. }

\label{fig:radioGasProfile}
\end{figure}

\subsection{Molecular gas masses}
\label{sec:masses}

An important property that can be derived from the ALMA data, are the masses
of the various kinematical components of the molecular gas. The integrated
total flux (after primary beam correction) corresponds to $76.4 \pm 3.8$ Jy
\kms. The flux of the gas in the region co-spatial with the radio is
$38.9 \pm 1.9$ Jy \kms of which $10.0 \pm 0.5$ Jy \kms are associated with gas
with anomalous velocities i.e.\ gas with observed velocities incompatible with
those of a regularly rotating disk with the rotation curve shown in Fig.\
\ref{modelRotCurve}, {i.e.\ gas rotating faster than expected and gas that
  shows apparent counter rotation}. The anomalous gas at the location of the W
lobe has a flux of about $2.7 \pm 0.3$ Jy \kms.

In order to convert these fluxes to masses of molecular gas, a number of
assumptions are needed.  As for the APEX data presented in
\citet{Morganti2013b}, we estimate the H$_2$ masses by assuming a luminosity
ratio $L^{\prime}$CO(2-1)/$L^{\prime}$CO(1-0) = 1, typical of thermalised and
optically thick gas, which could be the situation in the denser and warmer gas
associated with an outflow.

For the kinematically disturbed gas, we have considered a range of values for
the $X_{\rm CO}$ conversion factor.  This factor is assumed to be between $0.4
\times 10^{20}$ cm$^{-2}$ (K km s$^{-1}$)$^{-1}$ ($\alpha_{\rm CO} = 0.8$),
typical of a thick and dense medium and commonly used for ULIRG, and $0.2
\times 10^{20}$ cm$^{-2}$ (K km s$^{-1}$)$^{-1}$ (corresponding to
$\alpha_{\rm CO} = 0.34$, the most conservative case) derived for optically
thin cases and suggested for turbulent gas associated with winds (see the
discussion in \citet{Bolatto2013} and in \citet{Geach2014}).  Using these two
values for the CO-to-H$_2$ conversion factor, we derive masses for the
anomalous molecular gas ranging between $1.9$ and $4.8 \times 10^7$ \msun\ of
which between $0.5$ and $1.3 \times 10^7$ \msun\ is associated with the gas
with the most extreme kinematics at the location of the W lobe.  The result
is, therefore, that we confirm that the molecular outflow is more massive that
the \HI\ outflow which was estimated to be $3.6 \times 10^6$ \msun
\citep{Morganti2007}.  We also confirm the large difference between the mass
of the outflow of cold molecular gas, estimated here, and the mass of the warm
molecular gas estimated by \citet{Tadhunter2014}.

From the total integrated flux of the CO(2-1), we derive the total mass of the
molecular gas of the regularly rotating gas outside the region of the radio
plasma to be in the range $M_{\rm H_2} = 1.0 - 1.1 \times 10^9$ \msun\
assuming an $X_{\rm CO}$ conversion factor equal to $2 \times 10^{20}$
cm$^{-2}$ (K km s$^{-1}$)$^{-1}$ (corresponding to $\alpha_{\rm CO} = 4.6$,
typical for the Milky Way).  We consider the masses derived from the ALMA data
to be more reliable measurements that those obtained from the APEX data
\citep{Morganti2013b}. Two reasons should be considered. First of all, the
blueshifted wing of CO was covering most of the available band in the APEX
data. This made the definition of the continuum and the consequent continuum
subtraction very uncertain (as discussed already in
\citet{Morganti2013b}). The large observing band of the ALMA data has solved
this problem.  Furthermore, the distribution and kinematics of the anomalous
gas as observed in the ALMA data has turned out much more complex than was
assumed when measuring the gas outflow from APEX by \citet{Morganti2013b}.

\subsection{The radio continuum }
\label{sec:cont}

The continuum image at 230~GHz is shown in contours in Fig. \ref{fig:ContCO}
overlaid to the zoom-in of the central region of the CO(2-1). The morphology
of the continuum obtained from the ALMA data - a triple structure with the
core in the middle and two lobes of which the western one is the brightest -
is very similar to the structure observed at lower frequencies (17 and 24~GHz
with the ATCA) presented and discussed in \citet{Morganti2007}.  Using the
present and the previous radio continuum observations, we have derived the
spectral indices of the three components. The total fluxes of the three
components at 17 and 228~GHz are listed in Table{\ref{tab:radio}}, together
with the derived spectral indices. Between 17~GHz and 228~GHz we find a fairly
steep spectral index for the two lobes ($\alpha = 0.87$ for the E lobe and
$\alpha = 0.9$ for the W lobe; $S\propto \nu^{-\alpha}$) and a flatter spectral
index for the core ($\alpha = 0.33$). Thus the values obtained are similar,
although slightly steeper, than found between the much closer frequencies of
the ATCA data. As expected, these values support the non-thermal origin of the
emission with the power law extending to the high frequencies of ALMA. The
spectral indices also confirm the core and lobe classification of the three
structures.

\begin{table} 
  \caption{Flux densities and spectral indices for the radio three radio continuum components in IC~5063. }
\centering 
\begin{tabular}{lccccc} 
\hline\hline\\ 
Region &  17~GHz     &  228~GHz    & $\alpha$       \\
       &  mJy        &  mJy        &                \\
\hline
       &             &             &                 \\
   E   &      5.6    &      0.58   &  0.87           \\
  core &     30.8    &     12.8    &  0.33           \\
   W   &     91.2    &      6.2    &  0.93           \\
\hline
\end{tabular}
\label{tab:radio}
\end{table}

As mentioned above, the coincidence between the radio continuum and the
distribution of the bright molecular gas is clear. The CO(2-1) gas wraps
around the E lobe and part of the W lobe while the brighter CO part appears to
overlap (in projection) with the W hot-spot. It is interesting that the bright
CO region shows a strong correspondence with the radio continuum in the part
dominated by the most kinematically disturbed gas. As discussed in
Sec. \ref{sec:BrightInnerCO}, this can have influence on both the CO
excitation as well as production of CR.
 
It is also interesting to note that the radial distribution of the brightness
of CO and radio continuum shows an offset in the location of the peaks.  The
peaks of CO are at slightly larger distances from the centre compared to the
peaks of the radio continuum on both sides of the radio source, see Fig.\
\ref{fig:radioGasProfile}. We will comment on this further below, see Sec.\
\ref{sec:kin}.

\section{Discussion}

The observed distribution and kinematics of the cold molecular gas suggest
that the radio plasma jet is driving the bulk of the gas outflow.  Therefore,
we present a simple model which is aimed at explaining these characteristics
following the scenario presented in the numerical simulations of
\citet{Wagner2011} and \citet{Wagner2012}.  These simulations describe the
effects of a newly formed radio jet when moving though a {\sl dense clumpy
  medium}. A porous medium with dense clumps forces the jet to find the path
of least resistance, while interacting and gradually dispersing the dense
clouds away from the jet axis.  In this way, clouds can be accelerated to high
velocities and over a wide range of directions, and along the path of the jet
a turbulent cocoon of expanding gas forms, farther away from the jet axis.

Before describing the model, we first attempt to find clues to properties of 
the molecular gas in the observed kinematics as given in Fig.\ \ref{modelRotCurve}.

\begin{figure}
\centering
\includegraphics[width=\hsize, keepaspectratio]{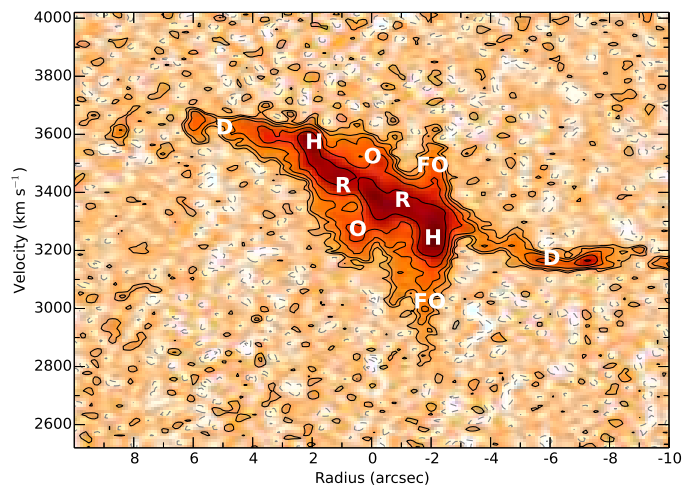}
\caption{Position-velocity diagram, as in Fig.\ \ref{modelRotCurve}, with the
  labels of various components identified (described in text). }
\label{sliceLabelled}
\end{figure}

\subsection{Signatures  of jet-ISM interaction in the  kinematics of the 
molecular gas in the inner region of IC 5063}
\label{sec:kin}

A closer look at Fig.\ \ref{modelRotCurve} shows that a number of key features
can be identified in the molecular gas. These are illustrated in Fig.\
\ref{sliceLabelled}. Beyond radii of $2^{\prime\prime}$ (i.e.\ beyond the
radio source), we detect the large-scale gas disk (labelled $D$), undisturbed
by the radio jet. At smaller radii, a bright S-shaped feature is seen. We
suggest that this bright feature actually consists of two structures. One of
these is a linear feature, of about 4 arcsec in size, which we labelled $R$ in
Fig.\ \ref{sliceLabelled}. When linear features are observed in $pv$ diagrams
of edge-on galaxies, such as IC 5063, they usually correspond to a rotating,
ringlike structure in the galaxy, or to a disk with solid-body rotation.
Solid-body rotation is not expected in the central regions of a massive
early-type galaxy such as IC 5063 \citep[e.g.][see also Fig.\
\ref{modelRotCurve}]{Cappellari2013}, so it is more likely that $R$
corresponds to a ring-like structure. However, Fig.\ \ref{modelRotCurve} shows
that if $R$ corresponds to a ring, its rotation velocity has to be lower than
the rotation speed expected from the mass model. We suggest that this may
indicate that feature $R$ corresponds to a rotating ring-like structure with a
radius of about $2^{\prime\prime}$ but rotating at velocities below nominal
rotation due to the interaction with the outflowing gas.

At both ends of $R$, there is bright emission (labelled $H$) near the location
of the two radio lobes, but the difference with $R$ is that they seem to
rotate at the expected rotation speed. Together with feature $R$, these give
the S-shape of the bright emission inside $r = 2^{\prime\prime}$.  This may
indicate that just outside the ring-like structure $R$ near the hot spots,
there is bright emission, but it is rotating at the nominal rotation speed.
This is also seen in Fig.\ \ref{fig:radioGasProfile} which shows that the
bright molecular gas at the locations of $H$ extends slightly beyond the radio
hotspots.  The excitation conditions of the gas just beyond the hot spots, as
seen from the centre, may already be changed by radiation from the jet-ISM
interaction but the kinematics has not been affected yet and the gas is still
rotating near the nominal rotation velocity.  The overall kinematics shows
that component $H$ only occurs near the radio lobes.

The other feature we identify is the fainter gas inside $r = 2^{\prime\prime}$
with velocities (very) inconsistent with the nominal rotation (component
$O$). In particular, part of this component has 'forbidden' velocities which
implies that the kinematics of this component is not dominated by rotation and
that it must involve a fast radial component. At the location of the W hot
spot the largest deviations from rotation are seen (component $FO$).

\begin{figure}
\centering
\includegraphics[width=\hsize, keepaspectratio]{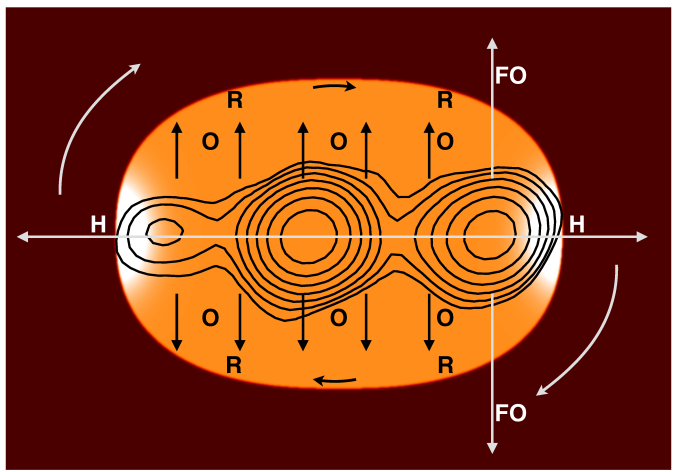}
\caption{Schematic illustration of the face-on view of the model described in 
Sec.\ \ref{sec:modeling}, with the contours of the continuum emission. }
\label{I5063Cartoon}
\end{figure}

\subsection{Modeling of the kinematics of the gas}
\label{sec:modeling}

Based on the discussion given above, we attempt to make a geometrical model of
the gas kinematics in IC~5063. The basic idea of the model is that a plasma
jet is expanding in the plane of the gas disk of IC 5063 and interacting with
this gas disk. This interaction is pushing the gas away from the centre and
creating a cocoon of laterally expanding gas along the radio jet.

To model the data cube, we employ a geometrical modelling technique often used
for modelling the kinematics of gas disks in galaxies
\citep[e.g.,][]{Jozsa2007}, i.e.\ we describe the gas disk in IC 5063 by a set
of concentric elliptical rings where each ring has an orientation, a density,
and a rotation- and radial velocity. We underline that we attempt to only
geometrically describe the kinematics of the gas; the model does not include
any physics to describe, e.g., the excitation of the gas or the actual
interaction of the jet plasma with the gas disk. For the inclination of the
rings, we choose 80$^\circ$ and for the position angle 115$^\circ$.

The face-on layout of the model is shown in Fig.\ \ref{I5063Cartoon} and
illustrates how IC 5063 would be seen from one of its poles. Along the radio
jet, the cocoon around the jet drives a lateral expansion (indicated by the
black vertical arrows) of the gas of the inner gas disk (light brown ellipse
in Fig.\ \ref{I5063Cartoon}) away from the jet axis. The interaction between
the expanding cocoon and the inner gas disk also causes the rotation velocity
in this region to be well below the nominal rotation, as we seem to observe in
Fig.\ \ref{modelRotCurve}.  This gas would correspond to component $O$. At the
radius where this laterally expanding gas meets the undisturbed outer disk, a
ring-like interface is formed of material with lower rotation velocities, but
the expansion is halted. We identify this with component $R$. The gas of the
quiescent, large-scale disk outside the zone of influence (dark brown Fig.\
\ref{I5063Cartoon}) is rotating according to the full rotation expected from
the mass distribution (Fig.\ \ref{modelRotCurve}).  The horizontal white arrow
Fig.\ \ref{I5063Cartoon} indicates the jet direction. The vertical white
arrows represent the extreme outflow velocities at the location of the W hot
spot due to the direct interaction of the jet with the ISM (component
$FO$). Just beyond the two hotspots, the rotation of the gas in the ISM is
following our mass model, but has higher brightness due to the nearby hotspot
(component $H$). We include this component also in order to reproduce the
observation that the molecular appears to extend somewhat beyond the bright
radio continuum source (see Fig.\ \ref{fig:radioGasProfile}).

We have attempted to reproduce the observed $pv$ diagram of IC~5063 by
constructing data cubes produced by models of the kind described above. Figure
\ref{modelRotCurve} shows that the kinematics and distribution of the gas is
not fully symmetric so any symmetric model will have difficulty in fitting the
data in detail. Therefore we have not attempted to replicate the fine details
of the brightness distribution and consequently the quality of a fit is not
assessed using a numerical goodness-of-fit criterion such, e.g., a
$\chi^2$. Instead, we have focussed on whether a model data cube reproduces
the {\sl locations} in the $pv$ diagram where gas is detected and where it is
not.

We have also attempted to keep the number of free parameters required to make
an acceptable match with the data, in the sense described in the previous
  paragraph to a minimum. For example, since we access the validity of a
model by a qualitative comparison with the data, the density/emissivity of the
gas of components $O$ and $R$ does not have to be a free parameter and is
assumed to be uniform (i.e.\ not to depend on position). In this way, the
model provides a better test for the assumed kinematics and geometry.

The main two parameters we found to be important to be able to describe the
data are the velocity of the lateral expansion of component $O$ and the
reduced rotation velocity of component $R$. Initially, we also varied the
orientation of the jet w.r.t.\ the line of sight. We found, however, this
parameter to be strongly degenerate with the outflow velocity of component $O$
in order to reproduce the observed (projected) velocities of component $O$,
i.e.\ for a jet oriented more in the direction of the line of sight, we have
to increase the lateral expansion velocity in order to obtain the same
projected velocity while not changing the qualitative aspects of the model
$pv$ diagrams. We have therefore chosen to consider only models for which the
jet is moving the plane of the sky; such models give a lower limit to the
expansion velocity of $O$. For simplicity, we have not included the
large-scale disk $D$ in our models.

We have found a good fit to the data using a model as described above for the
case that the lateral expansion reaches velocities up to 200 \kms\ and that
the rotation velocity of ring $R$ (and of component $O$) is $\sim$50 \kms\ at
$r = 2^{\prime\prime}$ (compared to a nominal rotation velocity of 170 \kms\
at that radius). Note that, since we have assumed the jet to move in the plane
of the sky, projection effects mean that the 200 \kms\ for the lateral outflow
is a lower limit. We have found it necessary, in order to reproduce the large
line widths observed for component $O$, to include that this component
exhibits a range of expansion velocities, from 0 \kms\ to 200 \kms. We have
chosen to implement this by letting the lateral expansion velocity depend
linearly on the lateral distance from the jet, but other choices, such as,
e.g., a turbulent medium with a large spread in velocity uncorrelated with
position, will give very similar results.

\begin{figure}
\centering
\includegraphics[width=\hsize, keepaspectratio]{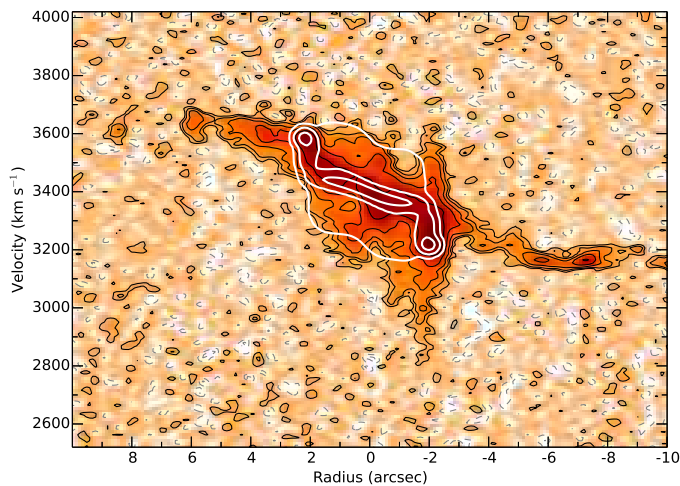}
\caption{Position-velocity diagram, as in Fig.\ \ref{modelRotCurve}, with
  white contours showing the model described in Sec.\ \ref{sec:modeling}. }
\label{modelSlice8}
\end{figure}

Figure \ref{modelSlice8} gives the observed $pv$ diagram with superposed on it
contours based on the above model. Given that we only seek qualitative
agreement with the data, we only draw those contours that indicate the
qualitative properties of the model.  Because we have assumed the spatial
brightness distribution to be uniform for $O$ and $R$, the brightness
distribution in the $pv$ diagram is the result of kinematics and geometry
only, not of any density/emissivity structure in the gas disk.  It is clear
that our model reproduces the S-shape of the bright inner feature very
well. Although the physical brightness of the affected regions $O$ and $R$ is
uniform in the model, the fact that $R$ forms a ring-like structure causes it
to project in the $pv$ diagram as a bright ridge. The bright regions just
outside the hot spots in our model reproduce $H$.  Although component $O$ has,
in the plane of the galaxy, the same brightness as $R$, it appears with lower
brightness in Fig.\ \ref{modelSlice8} because the same amount of emission is
spread over a much larger velocity range. The match for component $O$ is also
very acceptable, except for the extreme velocities near the western hot
spot($FO$).  Within the framework of our type of model, this can be alleviated
by choosing higher lateral expansion velocities at these locations. This is
justified because at this location the most intense jet-ISM interaction is
occurring, as evidenced by the very bright radio continuum emission from this
location. We suggest, similar to
\citet{Morganti1998,Oosterloo2000,Morganti2007,Tadhunter2014}, that at the W
hot spot, the jet is directly interacting with a cloud from the ISM.

We have considered several variations of the model given above. For example,
one can assume that the expansion of component $O$ is radial from the centre
instead of laterally from the jet. This would apply for a case where the
outflow is driven by the central AGN. We found, however, that such models do
not reproduce the observed $pv$ diagram very well. In particular, in such a
model, one would expect the observed anomalous velocities of component $O$ to
be largest near the centre and be close to zero near the hot spots. This is
quite different from observed. Another aspect we varied is the shape of the
region containing component $O$ and of the ring $R$, i.e.\ whether it is
circular or elliptical. Although this also depends on the radial density
distribution, we found that it is much easier to explain there is no obvious
observed radial trend in the brightness of $R$ and $O$ with models where $R$
and $O$ have an elliptical shape instead of circular. This can be roughly
understood by considering the extreme case of $R$ being a rectangle in which
case obviously there is no radial dependence introduced by the geometry.

\subsection{Hitting and Pushing: the two effects of the radio jet}
\label{sec:push}

The observed distribution and kinematics of the cold molecular gas, as well as
the model presented above, support the idea that the radio plasma jet is
driving the bulk of the gas outflow.  

In this respect, the model follows the idea at the base of the simulations of
\citet{Wagner2011} and \citet{Wagner2012}.  As mentioned above, these
simulations describe the effects of a newly formed radio jet when moving
though a dense clumpy medium.  One should note that, although the clouds do
initially have densities and temperatures that correspond to a molecular
phase, the simulations do not include the complete cooling of the gas and,
therefore, they do not follow exactly the same phase of the gas that we
observe.  However, although a direct comparison of the physical conditions of
the gas is not possible, the kinematics of the gas in the observations and
model presented here appears to compare remarkably well with the predictions
from the numerical simulations.

In our case, the clumpy medium is indeed traced by the bright CO around the
radio jet. As illustrated by, e.g., Fig.\ 4 in \citet{Wagner2011}, the
expanding, quasi-spherical energy bubble is responsible for pushing the clouds
outwards. Gas outflowing under
the effect of the expanding cocoon is expected to remain visible all along the
jet, in the same way as we have observed in IC~5063.

Interestingly, this causes the jet-ISM interaction to have an impact over a much
broader region than  of the direct cone of the jet. This is an important
element which makes this kind of interaction particularly relevant in the
context of feedback effects.

The kind of strong impact of a jet we observe would be expected when a newly formed
radio jet expands into the rich ISM surrounding a
young radio source. This has indeed been observed in ionized gas and \HI\ in
young radio sources \citep{Holt2009,Shih2013,Gereb2014b}. Although we do not
know whether IC~5063 is indeed a young radio source, the particular conditions
in IC~5063 where the jet is expanding in the plane of the large-scale gas disk
of the host galaxy, likely makes this object an ideal target to test this
scenario.

The velocities of the molecular gas observed with ALMA match those obtained by
the simulations for jets of comparable power to the one in IC~5063. In
addition to this, the velocity field expected in such models, as shown in
Fig.\ 4 of \citet{Wagner2011}, shows that the higher velocities are expected
at the location where the jet is interacting directly with the ISM, as is
happening in the W lobe. We claim that at this location we are seeing the {\sl
  direct} interaction of the jet with a large molecular cloud, which is
suggested to be present by the NICMOS results from \citet{Kulkarni1998}.  The
distribution of clouds in the surrounding ISM is likely not symmetrical near
the two lobes, which could explain why the situation in the E lobe is less
spectacular.

\subsection{The power source of the outflow}

In the preceding sections we have shown that a plasma jet expanding and
interacting in a gas-rich, clumpy medium appears to be a good explanation for
the morphology and kinematics of the gas as observed in our ALMA data. This
implies that the radio plasma jet is the main driver of the gas
outflow. However, IC~5063 has also a strong (optical) AGN and, therefore, we
now discuss the phenomena from an energy point of view.
 
Using the masses of the outflowing gas given in Sec.\ \ref{sec:masses}, we
derive the mass outflow rate using $\dot{M} = M/\tau_{\rm dyn}$ where $M$ is
the mass of the outflowing gas and $\tau_{\rm dyn}$ a timescale for the
outflow. Given the complex distribution of the outflowing gas, a number of
assumptions need to be made.  In particular, $\tau_{\rm dyn}$ was derived
using an intermediate distance from the core of 0.3 kpc and an average
velocity of the outflowing gas of 200 \kms, more typical of the gas along the
jet than the extreme values at the location of the W lobe.  This gives
$\tau_{\rm dyn} \simeq 1.5\times 10^6$ yr and we obtained $\dot{M}$ to lie in
the range 12 to 30 \msunyr, depending on the assumed 
CO-to-H$_2$ conversion factor.

The kinetic power of the outflowing gas can be estimated using the formula in
\citet{Holt2006} that includes, unlike the calculations in \citet{Cicone2014},
both the radial and turbulent components in the gas motion. Following
\citet{Holt2006}, we have assumed that the relatively large line-width of the
outflowing gas reflects a turbulent motion that is present at all locations in
the outflow region. Thus, the component of turbulent is represented by the
FWHM (full width at half maximum) of the lines:

\[
  \dot{E} = 6.34 \times 10^{35}\ \frac{\dot{M}}{2} \ \big(v^2_{\rm out} + \frac{v^2_{\rm turb}}{5.55}\big)\ 
\ \ \ \mathrm{erg\ s}^{-1}.
\]
where we have assumed $v_{\rm out}= 200$ \kms and
$v_{\rm turb} \sim FWHM \sim100$ \kms. These values are giving us a very
conservative estimate of the kinetic power of the outflow. Combining the \HI\
and CO components, this gives a kinetic power of the outflow of
$\sim 5 \times 10^{42}$ \ergs.

We can now compare this with the bolometric luminosity of the AGN and the jet
power.  The bolometric luminosity of the AGN is estimated to be
$L^{\rm bol}_{\rm AGN} = 7.7 \times 10^{44}$ \ergs\ \citep{Morganti2007}, 
thus the optical AGN is in principle capable of driving the outflow if the efficiency 
is higher than 0.6\%.

We estimate the jet power using the formulae given in \citet{Willot1999}, \citet{Wu2009}  and \citet{Cavagnolo2010}, and we obtain values between $5 \times 10^{43}$ \ergs\ and of $9 \times 10^{43}$ \ergs\ (using a conservative value of $f = 10$ for the normalisation uncertainty factor  introduced by \citealt{Willot1999}; see also \citealt{Wu2009} for more details). This indicates that the coupling between the energy transmitting to the medium by the jet and the ISM should have high efficiency (between 0.05 and 0.1) in order to drive the observed outflow.

It is interesting that both phenomena - AGN and radio jet - can power the
outflow. Despite of this, the observed kinematics suggests that the radio jet
plays the prominent role in driving the outflow.  This may indicate that
either it has a higher efficiency/coupling, or that perhaps both are at work
in the inner regions while the radio jet dominates at larger distances.
 
It is worth noting that IC~5063 was part of the sample included in the
analysis of \citet{Cicone2014}.  Interestingly, however, IC~5063 seems to
deviate from the correlation they found between the molecular outflow rate and
the AGN luminosity (their Fig.\ 9). This correlation was taken as a signature
of the optical AGN driving the outflow in the sample of local ULIRGs and QSO
hosts considered by these authors. The fact that IC 5063 does not lie on this
correlation indicates that the radio jet indeed could play a role in IC 5063.

\subsection{Interaction, shocks and cold gas}

Although the model described above can explain the kinematics of the gas quite
well, questions remain about how cold gas can be associated with such fast and
massive outflows, since shocks accelerate the gas and may heat it to $10^6 -
10^7$ K. Molecules should dissociate at such temperatures \citep{Hollenbach1989}. 
In the case of shocks produced by winds from the accretion disks,
\citet{Zubovas2014} have proposed that radiative cooling is fast enough to
reform molecules in a large fraction of the outflowing material on a short
time scale. They assume that, due to Rayleigh-Taylor instabilities, the cool
gas should collapse into clumps, which are then entrained in the hot-diffuse
medium. Key for the success of such a model is the time scale over which the
clumps can form.

A similar scenario, albeit with shocks driven by the radio plasma, has been
proposed for IC~5063 to explain the warm molecular gas kinematics
\citep{Tadhunter2014} and for the case of 4C~12.50 to explain the outflowing
\HI\ \citep{Morganti2013a}. In the latter study, a cooling time of the order
of $10^4$ years was derived for the outflowing cloud. Such a time scale is
comparable to the age of the radio source, suggesting that the gas has indeed
time to cool during the process of being expelled.  Finally,
\citet{Guillard2009,Guillard2012} proposed that the turbulent gas becomes
thermally unstable and cools as it moves. Therefore the molecular gas that
results from the post-shock cooling has the momentum of the tenuous gas that
has been accelerated to high velocities. This may explain why the molecular
gas is characterised by such high velocities.

In order to confirm such scenarios, the emissivity of gas strongly disturbed
by the passage of a radio jet and associated shocks needs to be followed with
simulations.  This has been investigated in a number of studies, although with
particular emphasis on young stellar and Herbig-Haro objects \citep[e.g.\
][]{Chernin1994,deGouveia1999,Raga2002}. Investigations, using a similar
approach, to trace the cooler gas by expanding to extragalactic sources the
simulations of relativistic jets presented in \citet{RochaDaSilva2015}, are
now in progress.  Preliminary results (Falceta-Gon{\c c}alves \& de Gouveia
Dal Pino priv.\ comm.) show that a relativistic jet entering a denser
environment is developing a turbulent cocoon. The consequent fast cooling,
associated with the turbulence (plus entrainment), results in a mix of gas
phases (from hot to cold) as is observed. Most importantly, the simulations
show, to first order, that the cooler gas is also characterized by high
velocities. Thus, despite the simplicity of the assumptions made (i.e.\
homogeneous ambient medium), the shocked ambient material deposited into the
cocoon around the jet includes a cool component (in addition to the warm and
hot) with large turbulent velocities, in agreement with the observations.

The results of our observations can also be used to investigate alternative
scenarios. For example, slow entrainment by the jet that would not dissociate
the molecules. However, if the warm H$_2$ is also heated during this
entrainment, the mass discrepancy between the warm H$_2$ ($M_{\rm H_2} \sim
8.2 \times 10^2$ $M_{\odot}$, \citet{Tadhunter2014}) and cooler component
derived from the CO(2-1) observations presented here, is difficult to explain.
Alternatively, if the heating of the clouds (and of the warm molecular gas) is
done by the AGN radiation field (i.e.\ in a similar way as seen in radiatively
excited molecular clouds, where the warm H$_2$ forms a skin around the colder
molecular material deeper in the clouds). However, our new CO(2-1)
observations make this possibility unlikely, because it is difficult to
explain how more than $10^7$ \msun\ of molecular gas (equivalent to an
extremely massive Giant Molecular Cloud) can be accelerated to the observed
high velocities without heating the molecules and destroying the clouds.

Thus, we conclude that the most likely explanation for the outflowing
molecular gas in IC~5063 is rapid post-shock cooling.

\section{Conclusions}

The ALMA data of IC~5063 presented here have provided a unique,
high-resolution 3-D view of the molecular gas in one of the best examples of
an on-going interaction between the radio jet and the ISM.  Because of its
proximity, and its geometry, this object is ideal to study the impact of radio
jets on the surrounding medium.  A wealth of data, including the warm
molecular gas, \HI\ and ionized gas, are already available for this
object. However, the ALMA data have provided a much more complete and complex
picture and have allowed to trace this interaction in much more detail (down
to scales of 0.5 arcsec, i.e.\ 100 pc) compared to what is available for the
other phases of the gas.  The results presented here further confirm the
multi-phase nature of AGN-driven outflows and the significant mass of the cold
gaseous component associated to it.

From these data, we have found kinematically disturbed gas distributed all
along the radio emission, with velocities strongly deviating from the
regularly rotating gas. This illustrates that gas outflows are produced not
only where the jet directly interacts with the ISM, but that the effects occur
over a much larger volume therefore making its impact larger, as already shown
by numerical simulations \citep[e.g.,][]{Wagner2011,Wagner2012}.
  
The region showing the disturbed kinematics of the gas is co-spatial with the
radio plasma and it also appears much brighter than the gas of the
larger-scale, regularly rotating component of molecular gas.  This inner,
brighter region of CO(2-1) is elongated along the radio axis and wrapping
around the radio continuum emission. The brightest CO(2-1) emission is
co-spatial with the bright, western radio lobe and has the largest velocities
deviating from regular rotation.  A simple model assuming lateral expansion of
the gas caused by the cocoon around the radio jet, describes the observed
kinematics remarkably well.

Taking all this together, we argue that the distribution and kinematics of the
molecular gas support the idea of the jet interacting with a clumpy medium and
pushing it outwards. In addition to the outflowing gas, this results in the
creation of a ring of gas, which represents the interface between the region
affected by the radio plasma and the undisturbed disk at larger radius. At
least at the location of the western hot spot, the jet is directly interacting
with a gas cloud, causing the very large outflow velocities observed. Such a
picture is fully consistent with the numerical simulations of
\citet{Wagner2011} and \citet{Wagner2012} of which all main components are
clearly identifiable in the data.

Even assuming the most conservative value of the conversion CO-to-H$_2$, we
find that the mass of the outflowing gas is between $1.8$ and
$4.5 \times 10^7$ \msun, thus much larger than the outflow observed in any
other gas phase. The mass outflow rates we derive are in the range 12 - 30
\msunyr.

Interestingly, looking at the energetics, it appears that both the bolometric
luminosity of the AGN, as well as the jet power, are large enough to drive the
outflow. The finding, based on the kinematics, that the radio plasma is
playing the dominant role (at least at radii larger than 100 pc), could
suggest a higher efficiency for this mechanism.

Particularly relevant from this study is the conclusion that the effect of the
radio plasma can be significant, also in objects, such as IC 5063, that are
often considered radio quiet. A number of cases where this situation may occur
have been recently identified (e.g.\ NGC~1266 \citep{Alatalo2011,Nyland2013};
NGC~1433 \citep{Combes2013}; J1430+1339 - nicknamed the ``Teacup AG'' -
\citep{Harrison2014}). However, direct evidence that the radio plasma is at
the origin of these outflows of cold gas is, so far, not available in all
these cases.  Another case is M51 where signatures of the interaction between
a weak radio jet and the ISM have been found in the molecular gas. A
high-density region traced by HCN has been detected along the weak radio jet
of this nearby object. This region shows a gradient in the HCN/CO ratio, with
decreasing values for larger distances from the jet
\citep{Matsushita2015}. This points to the effect of the jet pushing and
compressing the surrounding medium.  If the possibility of weak radio jets
driving massive gas outflows, due to their high coupling efficiency, is
confirmed for more objects, it would make even ``wimpy'' radio jets relevant
for the fate of gas in galaxies. Because such weak radio jets are more common
than the powerful ones, this can have implications for the role of feedback in
galaxy evolution.

Another interesting point in the context of feedback effects is the fate of
the kinematically disturbed gas. The observed outflow velocities of most of
the disturbed gas are only $\sim$200 \kms. Under such conditions, the gas will
not leave the galaxy, but most likely will just be relocated, in a process
comparable to a ``galactic fountain''.  Of the gas with disturbed kinematics,
only at most a very small fraction  may have velocities high enough to
leave the galaxy, a similar situation as observed in, e.g., NGC 1266
  \citep{Alatalo2015}.  However, the injection of the large amounts of energy
in the gas disk may have a similar effect as increasing the turbulence of the
gas, a process that is also considered to inhibit star formation \citep[see
the case of 3C326,][]{Guillard2015}.

In summary, the ALMA data presented here for IC~5063, show how spatially
resolved observations are key to being able to determine which mechanism,
radiation or jet, drives the observed kinetic feedback. Hopefully, expanding
this kind of work to more objects will finally allow us to quantify the impact
of an AGN on the ISM of the host galaxy.

\begin{acknowledgements}
  We thank Alex Wagner, Elisabete de Gouveia Dal Pino and Diego
  Falceta-Goncalves for their input and useful discussions.  This paper makes
  use of the following ALMA data: ADS/JAO.ALMA\#2012.1.00435.S. ALMA is a
  partnership of ESO (representing its member states), NSF (USA) and NINS
  (Japan), together with NRC (Canada) and NSC and ASIAA (Taiwan), in
  cooperation with the Republic of Chile. The Joint ALMA Observatory is
  operated by ESO, AUI/NRAO and NAOJ.  RM gratefully acknowledge support from
  the European Research Council under the European Union's Seventh Framework
  Programme (FP/2007-2013)/ERC Advanced Grant RADIOLIFE-320745.

\end{acknowledgements}


\begin{thebibliography}{}
\bibliographystyle{aa}

\bibitem[Alatalo et al.(2011)]{Alatalo2011} Alatalo K., et al. 2011, \apj, 735, 88 

\bibitem[Alatalo et al.(2013)]{Alatalo2013} Alatalo K., et al. 2013, \mnras, 432, 1796 

\bibitem[Alatalo et al.(2015)]{Alatalo2015} Alatalo, K., Lacy, M., 
Lanz, L., et al.\ 2015, \apj, 798, 31

\bibitem[Bayet et al.(2011)]{Bayet2011}Bayet et al. 2011,\mnras 414,1583

\bibitem[Best et al.(2005)]{Best2005} Best P.~N., Kauffmann G., Heckman T.~M., et al.\ 2005, \mnras, 362, 25 

\bibitem[Binney \& Tremaine(2008)]{Binney2008} Binney J., \& Tremaine S. 2008, Galactic Dynamics: Second Edition (Princeton University Press, Princeton)

\bibitem[Birzan et al.(2008)]{Birzan2008} B\^irzan et al. 2008, \apj, 686, 859; 

\bibitem[Bolatto et al.(2013)]{Bolatto2013} Bolatto A.~D., Wolfire M., \& Leroy A.~K. 2013, \araa, 51, 207 

\bibitem[Cappellari et al.(2013)]{Cappellari2013} Cappellari M., et al. 2013, \mnras, 432, 1709

\bibitem[Cavagnolo et al.(2010)]{Cavagnolo2010} Cavagnolo et al. 2010, \apj, 720, 1066

\bibitem[Chernin et al.(1994)]{Chernin1994} Chernin L., Masson C., Gouveia dal Pino E.~M., Benz W. 1994, \apj, 426, 204 

\bibitem[Combes et al.(2013)]{Combes2013} Combes F., et al. 2013, \aap, 558, A124

\bibitem[Combes et al.(2014)]{Combes2014} Combes F. 2014,  Proceedings of IAU Symp 309, eds B.L. Ziegler et al., arXiv:1408.1591

\bibitem[Cicone et al.(2014)]{Cicone2014} Cicone  et al. 2014, \aap, 562, A21 

\bibitem[Colina et al.(1991)]{Colina1991} Colina L., Sparks W.~B., Macchetto F. 1991, \apj, 370, 102 

\bibitem[Costa et al.(2014)]{Costa2014} Costa T., Sijacki D., Haehnelt M.~G. 2014 \mnras, 444, 2355 

\bibitem[Dasyra \& Combes(2011)]{Dasyra2011} Dasyra K.~M., Combes F. 2011, \aap, 533, L10 

\bibitem[Dasyra \& Combes(2012)]{Dasyra2012} Dasyra K.~M., Combes F. 2012, \aap, 541, L7

\bibitem[Davis et al.(2013)]{Davis2013} Davis T.~A., et al. 2013, \mnras, 429, 534 

\bibitem[de Gouveia Dal Pino(1999)]{deGouveia1999} de Gouveia Dal Pino E.~M., 1999, \apj, 526, 862 

\bibitem[Fabian(2013)]{Fabian2013} Fabian, A. 2013,  \araa, 50, 455

\bibitem[Faucher-Gigu{\`e}re \& Quataert(2012)]{Faucher2012} Faucher-Gigu{\`e}re C.-A., Quataert E. 2012, \mnras, 425, 605 

\bibitem[Ferland et al.(2009)]{Ferland2009}Ferland et al. 2009, \mnras 392, 1475
 
\bibitem[Feruglio et al.(2010)]{Feruglio2010} Feruglio C., Maiolino R., Piconcelli E., Menci N., Aussel H., Lamastra A., Fiore F. 2010, \aap, 518, L155 

\bibitem[Garc{\'{\i}}a-Burillo et al.(2014)]{Garcia2014} Garc{\'{\i}}a-Burillo S. et al. 2014, \aap, 567, 125 

\bibitem[Geach et al.(2014)]{Geach2014} Geach J.~E., et al. 2014, \nat, 516, 68 

\bibitem[Gereb et al.(2014a)]{Gereb2014a} Ger{\'e}b K., Morganti R., Oosterloo T.~A. 2014a, \aap, 569, AA35

\bibitem[Gereb et al.(2014b)]{Gereb2014b} Ger{\'e}b K., Maccagni F., Morganti  R., Oosterloo T.  2014b, \aap, in press., arXiv:1411.0361

\bibitem[Guillard et al.(2009)]{Guillard2009} Guillard P., Boulanger F., Pineau Des For{\^e}ts G., Appleton P.~N., 2009, \aap, 502, 515 

\bibitem[Guillard et al.(2012)]{Guillard2012} Guillard P., et al., 2012, ApJ, 747, 95 


\bibitem[Guillard et al.(2015)]{Guillard2015} Guillard P., et al. 2015, \aap, 547, 32


\bibitem[Harrison et al.(2014)]{Harrison2014} Harrison C.~M. et al. 2014, \apj, submitted, arXiv:1410.4198

\bibitem[Hollenbach \& McKee(1989)]{Hollenbach1989} Hollenbach, D., \& McKee, C.~F.\ 1989, \apj, 342, 306 

\bibitem[Holt et al.(2006)]{Holt2006} Holt J., Tadhunter C., Morganti R., Bellamy M., Gonz{\'a}lez Delgado R.~M., Tzioumis A., Inskip K.~J. 2006, \mnras, 370, 1633

\bibitem[Holt et al.(2009)]{Holt2009} Holt J., Tadhunter C.~N., Morganti R. 2009, \mnras, 400, 589 

\bibitem[Loenen et al.(2008)]{Loenen2008} Loenen et al.\ 2008, \aap, 488, 5L

\bibitem[J{\'o}zsa et al.(2007)]{Jozsa2007} J{\'o}zsa G.~I.~G., Kenn F., Klein U., Oosterloo T.~A. 2007, \aap, 468, 731 

\bibitem[Kanekar \& Chengalur(2008)]{Kanekar2008} Kanekar, N. \& Chengalur, J.~N. 2008, \mnras, 384, 6L

\bibitem[Krause et al.(2007)]{Krause2007} Krause M., Fendt C., Neininger N. 2007, \aap, 467, 1037  

\bibitem[Kulkarni et al.(1998)]{Kulkarni1998} Kulkarni, V.P., Calzetti, D.,  Bergeron, L. et al.\ 1998, \apj, 492, L121  

\bibitem[Mahony et al.(2013)]{Mahony2013}Mahony E., Morganti R., Emonts B., Oosterloo T., Tadhunter C. 
2013, \mnras, 435, L58

\bibitem[Maloney et al.(1996)]{Maloney1996}Maloney et a. 1996, \apj 466, 561

\bibitem[Martin et al.(1989)]{Martin1989}Martin, P., Roy, J-R., Noreau, L., LO, K.Y. 1989, \aap, 345, 707

\bibitem[Matsushita et al.(2007)]{Matsushita2007} Matsushita, S., Muller, S., Lim, J. 2007, \aap, 468, L49

\bibitem[Matsushita et al.(2015)]{Matsushita2015} Matsushita S., Trung D.-V.-, Boone F., Krips M., Lim J., Muller S. 2015, \apj, 799, 26


\bibitem[McMulline et al.(2007)]{McMullin2007} McMullin J.~P., Waters B., Schiebel D., Young W., Golap K., 2007, ASPC, 376, 127 

\bibitem[McNamara \& Nulsen(2012)]{McNamara2012} McNamara B.~R., Nulsen P.~E.~J., 2012, NJPh, 14, 055023 

\bibitem[Meijerink et al.(2007)]{Meijerink2007}Meijerink et al. 2007, \aap 461, 793
\bibitem[Meijerink et al.(2011)]{Meijerink2011}Meijerink et al. 2011,  \aap 525, 119

\bibitem[Morganti et al.(1998)]{Morganti1998}Morganti, R., Oosterloo, T.A., Tsvetanov, Z. 1998, \aj, 115, 915

\bibitem[Morganti et al.(2005a)]{Morganti2005a}Morganti, R., Tadhunter, C. N., Oosterloo, T. A. 2005a, \aap, 444, L9 

\bibitem[Morganti et al.(2005b)]{Morganti2005b}Morganti et al. 2005b, \aap, 439, 521; 

\bibitem[Morganti et al.(2007)]{Morganti2007} Morganti R., Holt J., Saripalli L., Oosterloo T.~A., 
Tadhunter C.~N. 2007, \aap, 476, 735


\bibitem[Morganti et al.(2013a)]{Morganti2013a}  Morganti, R., Fogasy, J., Paragi, Z., Oosterloo, T., 
Orienti, M. 2013a,  Sci, 341, 1082

\bibitem[Morganti et al.(2013b)]{Morganti2013b} Morganti R., Frieswijk W., Oonk R., Oosterloo T., Tadhunter C. 2013b,  \aap,  552, L4

\bibitem[Morganti(2014)]{Morganti2014}  Morganti R. 2014, in Extragalactic jets from every angle, Galapagos, ed. F. Massaro, C. C. Cheung, E. Lopez, and A. Siemiginowska (Cambridge University Press),  arXiv:1411.6107

\bibitem[Nyland et al.(2013)]{Nyland2013} Nyland, K., et al.\ 2013, \apj, 779, 173
 
\bibitem[Oosterloo et al.(2000)]{Oosterloo2000} Oosterloo T., Morganti R., Tzioumis A.,  Reynolds J., King E. et al. 2000, \aj, 119, 2085 

\bibitem[Punsly(2005)]{Punsly2005} Punsly, B. 2005, \apj, 623, L9

\bibitem[Raga et al.(2002)]{Raga2002} Raga A.~C., de Gouveia Dal Pino E.~M., Noriega-Crespo A., Mininni P.~D., Vel{\'a}zquez P.~F. 2002, \aap, 392, 267 

\bibitem[Rocha da Silva(2015)]{RochaDaSilva2015} Rocha da Silva, G., Falceta-Gon{\c c}alves, D., Kowal, G., \& de Gouveia Dal Pino, E.~M.\ 2015, \mnras, 446, 104

\bibitem[Sault et al.(1995)]{Sault1995} Sault R. J., Teuben P. J., Wright M. C. H., 1995, ASPC, 77, 433

\bibitem[Serra et al.(2012)]{Serra2012} Serra P., et al. 2012, \mnras, 422, 1835 

\bibitem[Serra et al.(2014)]{Serra2014} Serra P., et al. 2014, \mnras, 444, 3388 


\bibitem[Shih et al.(2013)]{Shih2013} Shih H.-Y., Stockton A., Kewley L. 2013, \apj, 772, 138

\bibitem[Sternberg et al.(1998)]{Sternberg1989}Sternberg et al. 1989,  \apj 320, 676

\bibitem[Tadhunter(2008)]{Tadhunter2008} Tadhunter C. 2008, MmSAI, 79, 1205 

\bibitem[Tadhunter et al.(2014)]{Tadhunter2014} Tadhunter C., Morganti R., Rose M., Oonk J.~B.~R., Oosterloo T. 2014, \nat, 511, 440;

\bibitem[Tielens \& Hollenbach(1985)]{Tielens1985} Tielens \& Hollenbach 1985, \apj 291, 722

\bibitem[van Driel \& van Woerden(1991)]{vanDriel1991} van Driel W., van Woerden H. 1991, \aap, 243, 71

\bibitem[Wagner \& Bicknell(2011)]{Wagner2011} Wagner A.~Y., Bicknell G.~V. 2011, \apj, 728, 29 

\bibitem[Wagner et al.(2012)]{Wagner2012} Wagner A.~Y., Bicknell G.~V., Umemura M. 2012, \apj, 757, 136 

\bibitem[Wiklind et al.(1995)]{Wiklind1995} Wiklind T., Combes F., Henkel C. 1995, \aap, 297, 643

\bibitem[Willot et al.(1999)]{Willot1999}Willott, C., Rawlings, S., Blundell, K., Lacy, M. 1999, \mnras, 309, 1017

\bibitem[Wright(2006)]{Wright2006} Wright E.~L. 2006, \pasp, 118, 1711 

\bibitem[Wu(2009)]{Wu2009} Wu Q. 2009, \mnras, 398, 1905

\bibitem[Zubovas \& King(2012)]{Zubovas2012} Zubovas K., King A. 2012, \apj, 745, L34 

\bibitem[Zubovas \& King(2014)]{Zubovas2014} Zubovas K., King A. 2014, \mnras, 439, 400 

\bibitem[Young et al.(2011)]{Young2011}  Young L.~M., et al. 2011, \mnras, 414, 940 
\end{thebibliography}
\end{document}